\documentclass[preprint,nofootinbib,noshowkeys,noshowpacs,longbibliography,
superscriptaddress,tightenlines,floatfix]{revtex4-2}
\usepackage[colorlinks=true,linktocpage=true,linkcolor=darkblue,citecolor=red,urlcolor=darkblue]{hyperref}
\usepackage{amsmath,amsfonts,amsthm,amssymb}
\usepackage{pifont}
\usepackage{graphics,graphicx}
\usepackage{color}
\usepackage{orcidlink}
\definecolor{darkblue}{RGB}{0,0,196}
\definecolor{darkgreen}{RGB}{0,120,0}

\def\beq{\begin{eqnarray}}
\def\eeq{\end{eqnarray}}

\newcommand{\C}{\mathcal C}
\newcommand{\vol}{\mathrm{vol}}
\newcommand{\dd}{\mathrm d}
\newcommand{\CP}{\mathsf{CP}}

\begin{document}
\preprint{}
 
\title{Anomalous charge conjugation and parity in \texorpdfstring{$(2+1)$}{}-dimensional relativistic hydrodynamics}
\author{Sonali Patnaik\orcidlink{0000-0003-4293-3725}}
\email{spatnaik@iitg.ac.in}
\affiliation{Indian Institute of Technology, North Guwahati, Guwahati 781039, Assam, India}
\author{Rajeev Singh\orcidlink{0000-0001-5855-4039}}
\email{rsingh@itp.uni-frankfurt.de}
\affiliation{Institut f$\ddot{u}$r Theoretische Physik, Goethe University,
Max-von-Laue-Straße 1, D-60438 Frankfurt am Main, Germany}
\date{\today} 
\bigskip
\bigskip
\begin{abstract}
We present a hydrostatic and first-order constitutive description of a mixed anomaly that involves $U(1)_Q$, $U(1)_F$, charge conjugation symmetry (acting only on $U(1)_Q$), and parity symmetry in $2+1$ dimensions. It is represented by a mutual Chern--Simons level $k_{QF}=\frac12 \,(\mathrm{mod}\ \mathbb Z)$. Adopting the already established multi-charge anomaly sector, we explicitly provide the information that this response is \textit{odd} under $\mathsf{C}$ and $\mathsf{P}$ symmetries separately, but \textit{even} under combined $\mathsf{CP}$ symmetry. We also provide a global inflow representative and compute the equilibrium cross-charge response which includes $\partial n_Q/\partial B_F=\partial n_F/\partial B_Q=1/4\pi \bmod 1/2\pi$. Using an analytic neutral-fluid example we show that this response is non-vanishing whereas the source-free hydrodynamic poles are invariant.
\end{abstract}
\maketitle
\newpage
\tableofcontents
\newpage
\section{Introduction}
\label{sec:intro}
Quantum anomalies constrain the long wavelength description of a system even when its microscopic degrees of freedom no longer provide a useful description of its dynamics. In a fluid, these constraints can arise in constitutive relations and equilibrium response~\cite{Son:2009tf,Jensen:2012jy,Amoretti:2022vxq,Amoretti:2023hpb,Martinoia:2024hip}. The study done by Son and Sur\'owka provides that triangle anomalies, together with the local second law, require non-dissipative contributions to relativistic charge currents~\cite{Son:2009tf}. However, equilibrium generating functionals provided another way to obtain such constraints and to distinguish thermodynamic response from transport information that vanish in stationary profiles~\cite{Banerjee:2012iz,Jensen:2012jh}. It was also noted that anomaly inflow provides the connection to the underlying quantum theory: an anomalous variation in a given spacetime dimension is compensated by a bulk contribution in one higher dimension~\cite{Callan:1984sa}. Its implementation in thermal equilibrium and in effective actions for fluids with anomalies give a systematic road-map from anomaly data to hydrodynamic constitutive relations~\cite{Jensen:2013kka,Haehl:2013hoa,Jensen:2013rga}.

This is particularly true for discrete anomalies, which need to be handled with extra care. They do not lead to an anomalous divergence in the local current Ward identity, and their effects generally cannot be inferred solely from local conservation relations. The parity anomaly in $2+1$ dimensions provides an intuitive example, and the initial works gave the relation between fermion regularization, gauge invariance, and parity-odd terms in the effective gauge action~\cite{Niemi:1983rq,Redlich:1983dv}. In addition, the physical realization in the condensed matter system was provided by Haldane’s lattice model, where the quantized Hall response appears without an external magnetic field and connects its critical fermionic spectrum to the parity anomaly~\cite{Haldane:1988zza}. The bulk perspective makes the global obstruction manifest and relates it to the boundary description of a topological phase~\cite{Witten:2015aba,Witten:2019bou}. The topological field theory of time-reversal-invariant insulators links bulk electromagnetic response to boundary Hall physics via dimensional reduction~\cite{Qi:2008ew}. In the language of correlation functions, the background Chern-Simons terms give rise to a parity-odd contact contribution. Allowed local counterterms can change their integer parts, however, the fractional parts carry information that cannot be removed by this freedom~\cite{Closset:2012vp}. Thus, a theory with anomalies may follow ordinary hydrodynamic conservation laws but may retain a non-trivial dependence on the global structure of its sources~\cite{Hidaka:2017auj}.

The hydrodynamics of parity-violating normal fluids in $(2+1)$D was presented at first order in gradients in Refs.~\cite{Jensen:2011xb,Chen:2013dca}. Their analysis distinguishes several non-dissipative coefficients and connects part of this sector with the equilibrium response to magnetic fields and vorticity. More broadly, thermal effective theories show how global anomalies constrain Chern-Simons couplings, and hydrodynamic effective actions clarify the restrictions imposed by discrete symmetries~\cite{Golkar:2015oxw,Glorioso:2017lcn}. These developments also make clear why parity-odd response and anomaly induced response should not be identified without further assumptions: symmetry allows for constitutive information whose values are not specified by a given anomaly.

Our study is particularly inspired by two studies done before. First, Ref.~\cite{Poovuttikul:2021mxx} studied anomaly inflow with the equilibrium partition function to obtain the parity anomaly contribution to first-order hydrodynamics in $2+1$ dimensions. And, second, Ref.~\cite{Furusawa:2021vlw} constructed the associated identically conserved bulk current and computed the charge and entropy currents at a physical boundary. Their treatment includes more than one $U(1)$ symmetries and explicitly discusses an off-diagonal two-charge anomaly matrix proportional to $\sigma_x/(4\pi)$. Furthermore, mixed symmetry anomalies are studied in~\cite{Wang:2017txt} for the description of deconfined quantum critical points. The study of this mixed anomaly sector and the corresponding multi-charge edge-current relations are important ingredients of our current work. Our focus is instead on a specific realization of this structure in which the discrete transformation acts as charge conjugation on one $U(1)$ factor while leaving the second $U(1)$ factor unaffected, and on tracing its consequences systematically through relativistic hydrodynamics and source response.

We consider a normal relativistic fluid in $2+1$D with compact global symmetry $G = \left[ U(1)_Q \rtimes Z_2^C \right]\times U(1)_F$ coupled to non-dynamical background connections $A_Q$ and $A_F$. The internal involution denoted by $\mathsf C$ reverses the $Q$ charge and leaves the $F$ charge invariant. Parity $\mathsf P$ reflects one spatial coordinate and acts on both the connections as ordinary one-forms. This choice of symmetry action is important. For a single ordinary $U(1)$ connection with $A\rightarrow-A$, the characteristic form $F\wedge F$ is even under $\mathsf C$. In contrast, with two charge sectors the mixed characteristic form $F_Q\wedge F_F$ is odd under $\mathsf C$. Its non-trivial quantized bulk phase gives a shared obstruction to preserving $\mathsf C$ and $\mathsf P$ individually in an intrinsic three-dimensional regularization that preserves the compact background gauge symmetries. The selected local response preserves $\mathsf{CP}$. We note that throughout this work, anomalous charge conjugation and parity refer to this specified approach, with one mixed anomaly coefficient.

Our aim is to follow this anomaly through the global source construction, hydrostatic functional, hydrodynamic frame matching, and observable source response within a single analytic model. We retain the compact charge lattice, since the separation between an admissible counterterm and a change of anomaly class depends on the normalization of both the charges. We also specify the bulk phase in a charge conjugation background and evaluate a thermal mapping torus example. The hydrodynamic description itself assumes a locally equilibrated phase with two conserved charge densities and no extra slow variables. This phase is an input to the construction and its existence is not obtained from the anomaly.

In equilibrium, the transgression of the mixed bulk term determines the magnetic and vortical contributions associated with the chosen anomaly. A direct variation of the local mutual Chern-Simons functional provides an independent check. In this representation, the anomalous currents depend solely on the background field strengths, while the local bulk stress contribution vanishes, and the two Lorentz force contributions cancel. Matching the energy density and both charge densities then gives the corresponding Landau frame constitutive relations, including the magnetic pressure terms determined by the thermodynamic derivatives. We also assume a parity preserving regular hydrostatic sector and provide the additional constitutive choices away from equilibrium. The resulting anomaly contribution is compatible with the local second law, however, the equation of state, dissipative coefficients, and optional non-hydrostatic Hall terms remain separate inputs.

This construction also allows a direct comparison between equilibrium contact response and collective dynamics. We obtain reciprocal magnetic density responses and the associated transverse static source kernel, keeping its order of limits manifest. The resulting response is a cross-charge response: a background field in one $U(1)$ sector induces a transverse response in the other sector, with its fractional part fixed by the anomaly class. Equilibrium magnetization currents and a full homogeneous DC conductivity probe different information, so the fractional local coefficient is not identified with an otherwise unrestricted transport measurement~\cite{Cooper_1997}. In an analytic example near a neutral equilibrium, we show that the cross charge response remain non-vanishing while the source free sound, shear, and charge diffusion poles are unaffected by the mixed coefficient. The example shows how the anomaly can be seen in local response without incorporating an additional hydrodynamic mode.

Finally, a physical edge requires information beyond the interior constitutive relations. We use the multi-charge edge current construction of Ref.~\cite{Furusawa:2021vlw}, with its boundary ansatz and the undetermined thermal coefficient mentioned explicitly. For an edge geometry and constitutive sector that preserves $\mathsf {CP}$ symmetry, we examine the additional restriction on that coefficient. This restriction follows from the boundary symmetry assumption, and it is not a general determination by the mixed gauge anomaly. Keeping the physical edge separate from the auxiliary boundary used for inflow also clarifies how bulk contact terms, magnetization currents, and edge currents must be consistently assigned.

Section~\ref{sec:setup} presents the symmetry setup and specifies the hydrodynamic assumptions followed by the discussion on mixed inflow, the counterterm obstruction, and the global background test in Sec.~\ref{sec:inflow}. Section~\ref{sec:hydrostatic} computes the hydrostatic functional by transgression, whereas, Sec.~\ref{sec:constitutive} presents the constitutive relations, performs the two-charge Landau-frame matching, and discusses entropy production. Static response and the analytic zero-field example are given in Sec.~\ref{sec:response} and Sec.~\ref{sec:edges} addresses physical edges and the limits of bulk anomaly matching. Finally, we give the summary of our work in Sec.~\ref{sec:summary}.
\section{Symmetry setup and assumptions of hydrodynamics}
\label{sec:setup}
We work with a normal relativistic fluid in $2+1$D with two compact, unbroken global symmetries $U(1)_Q$ and $U(1)_F$. Their external connections are given as $A_Q$ and $A_F$, with curvatures as
\begin{equation}
F_a=\dd A_a,
\qquad
\int_{\Sigma_2}\frac{F_a}{2\pi}\in\mathbb Z,
\qquad {\rm where} \qquad a=Q,F,
\end{equation}
for every closed two-cycle $\Sigma_2$. The unit-charge lattice is a part of the
definition. Neither connection is dynamical. We assume a
locally equilibrated phase whose only bulk hydrodynamic variables are the temperature, $T$, and the fluid velocity, $u^\mu$, and the two chemical potentials $\mu_a$.
Additional Goldstone fields, approximately conserved densities, and
independent topological sectors are not included in this constitutive model.

The internal involution discrete transformation called charge conjugation is
\begin{equation}
 \mathsf C:(A_Q,A_F)\longmapsto(-A_Q,A_F),
 \qquad (\mu_Q,\mu_F)\longmapsto(-\mu_Q,\mu_F).
 \label{eq:Caction}
\end{equation}
Thus $\mathsf C$ conjugates the $Q$ charge and leaves the $F$ charge invariant. The corresponding global symmetry is~\cite{Wang:2017txt}.
\begin{equation}
 G =
 \left[U(1)_Q\rtimes\mathbb Z_2^{\mathsf C}\right]\times U(1)_F ,
 \label{eq:symmetry_group}
\end{equation} 
where $\mathbb Z_2^{\mathsf C}$ acts on the $U(1)_Q$ by the inversion automorphism $A_Q\mapsto-A_Q$ and acts trivially on the $U(1)_F$.
In particular, $\mathsf C$ should not be seen as an operation that reverses both $U(1)$ charges. This asymmetric action is important for the mixed construction considered below: while $F_Q\wedge F_Q$ is \textit{even} under $\mathsf C$, the mixed form $F_Q\wedge F_F$ is \textit{odd}.

As mentioned in the introduction, the parity $\mathsf P$ reflects one spatial coordinate, and both connections transform as ordinary one-forms. In particular, both the chemical potentials are parity even. We take $\mathsf C$ and $\mathsf P$ to commute, so that the combined transformation $\mathsf{CP}$ is also a symmetry of the selected anomaly response. The mixed anomaly specified below obstructs the preservation of either operation separately in an intrinsic 3D regularization preserving the compact background gauge symmetries, while the combined $\mathsf{CP}$ transformation leaves the corresponding response unaffected. It is one shared mixed anomaly, not two independently adjustable anomaly coefficients. It is useful to summarize the transformation properties of the basic hydrodynamic variables and the first-order hydrostatic structures, see Table~\ref{tab:symmetry_transformations}.
\begin{table}[t]
\centering
\renewcommand{\arraystretch}{1.15}
\begin{tabular}{c|ccc}
\hline\hline
Quantity & $\mathsf C$ & $\mathsf P$ & $\mathsf{CP}$ \\
\hline
$T$              & $+$ & $+$ & $+$ \\
$\mu_Q$          & $-$ & $+$ & $-$ \\
$\mu_F$          & $+$ & $+$ & $+$ \\
$n_Q$            & $-$ & $+$ & $-$ \\
$n_F$            & $+$ & $+$ & $+$ \\
$B_Q$            & $-$ & $-$ & $+$ \\
$B_F$            & $+$ & $-$ & $-$ \\
$\Omega$         & $+$ & $-$ & $-$ \\
\hline
$\mu_F B_Q$      & $-$ & $-$ & $+$ \\
$\mu_Q B_F$      & $-$ & $-$ & $+$ \\
$\mu_Q\mu_F\Omega$ & $-$ & $-$ & $+$ \\
\hline\hline
\end{tabular}
\caption{Transformation properties of the thermodynamic variables and first-order hydrostatic structures under the discrete transformations specified in Eq.~\eqref{eq:Caction}. The three structures in the last three rows are odd under $\mathsf C$ and $\mathsf P$ separately, but even under $\mathsf{CP}$.}
\label{tab:symmetry_transformations}
\end{table}

We use the `mostly-plus' metric signature $(-,+,+)$ with fluid-velocity normalized as $u^\mu u_\mu=-1$. The Levi-Civita convention follow $\epsilon^{012}=1/\sqrt{-g}$. On an oriented coordinate patch, we set $\vol_3=\sqrt{-g}\,\dd x^0\wedge\dd x^1\wedge\dd x^2$. Our definitions are
\begin{align}
 \Delta^{\mu\nu}&=g^{\mu\nu}+u^\mu u^\nu,
 & E_a^\mu&=F_a^{\mu\nu}u_\nu,
 & B_a&=-\frac12\epsilon^{\mu\nu\rho}u_\mu F_{a\nu\rho},\nonumber\\
 \omega_{\mu\nu}&=\Delta_\mu{}^\alpha\Delta_\nu{}^\beta
\nabla_{[\alpha}u_{\beta]},
 & \Omega&=-\epsilon^{\mu\nu\rho}u_\mu\omega_{\nu\rho},
 & \widetilde X^\mu&=\epsilon^{\mu\nu\rho}u_\nu X_\rho .
 \label{eq:conventions}
\end{align}
The convention for anti-symmetrization includes a factor of $1/2$. 
With $\omega=\frac12\omega_{\mu\nu}\dd x^\mu\wedge\dd x^\nu$, these conventions imply 
\beq
u\wedge F_a &=& -B_a \, \vol_3 \,,\nonumber \\
u\wedge\omega &=& -\Omega\, \vol_3/2.
\eeq
We use the weak-field derivative counting, where $F_a=O(\partial)$ is of the order of gradients and
$\nabla T,\nabla\mu_a,\nabla u=O(\partial)$ is also of first order in gradients. But the fields $T$ and $\mu_a$ are of order $O(1)$. The equation of state follows the thermodynamic relations
\begin{equation}
 \dd p=s\,\dd T+n_a\,\dd\mu_a,
 \qquad \epsilon+p=Ts+\mu_an_a\equiv w,
 \qquad \dd\epsilon=T\,\dd s+\mu_a\,\dd n_a.
 \label{eq:thermo}
\end{equation}
Note the repeated charge labels are summed. We work at $T>0$, $w>0$, away from thermodynamic singularities, and require the map $(T,\mu_Q,\mu_F)\mapsto(\epsilon,n_Q,n_F)$ to be locally invertible. At zero field, unbroken $\mathsf C$ requires
$p(T,\mu_Q,\mu_F)=p(T,-\mu_Q,\mu_F)$.

The local Ward identities in the interior are
\begin{equation}
 \nabla_\mu J_a^\mu=0,
 \qquad \nabla_\mu T^{\mu\nu}=F_a^{\nu\lambda}J_{a\lambda}.
 \label{eq:wards}
\end{equation}
There is no perturbative non-conservation of either $U(1)$ current in the bulk. The discrete anomaly is instead present in the global dependence of the generating functional on the background sources and in the fractional part of the parity-odd contact response. Thus, the anomaly considered here does not modify the conservation laws.

The hydrostatic construction follows the inflow approach of Ref.~\cite{Poovuttikul:2021mxx}. Multi-charge discrete-anomaly hydrodynamics, including an off-diagonal anomaly matrix proportional to $\sigma_x/(4\pi)$, was presented in Ref.~\cite{Furusawa:2021vlw}. We take the corresponding multi-charge anomaly sector as an input. Our focus is on the specific symmetry realization in Eq.~\eqref{eq:symmetry_group}, in which the $\mathbb Z_2^{\mathsf C}$ transformation reverses one charge and leaves the other unchanged, and on following the resulting anomaly representative through hydrostatic transgression, constitutive matching, static cross-charge response, and collective hydrodynamics.
\section{Mixed inflow and the counterterm obstruction}
\label{sec:inflow}
\subsection{Why one ordinary charge is insufficient in this construction}
The specific discrete anomaly considered here requires a characteristic form that is odd under the charge-conjugation action defined in Eq.~\eqref{eq:Caction}. For a single ordinary $U(1)$ connection with $A\mapsto-A$, the smooth four-form $F\wedge F$ is charge-conjugation even. Pure gravitational characteristic forms are also even under this internal transformation. Hence, on a torsion-free background with no additional gauge or higher-form sources, these terms cannot provide the charge-conjugation-odd local inflow sought here. This is a statement about the characteristic-form mechanism and its first-order local descendant. The second compact $U(1)$ allows the mixed characteristic form
\begin{equation}
 I_4=\frac{F_Q}{2\pi}\wedge\frac{F_F}{2\pi},
 \qquad \int_{X_4}I_4\in\mathbb Z\,,
 \label{eq:I4}
\end{equation}
on a closed oriented four-manifold $X_4$. With the parity transformation adopted in this work, the same four-form is odd under $\mathsf P$ and $\mathsf C$ symmetries. Consequently, the phase $\exp(i\vartheta\int I_4)$ is invariant under each operation at $\vartheta=0$ or $\pi$ modulo $2\pi$. We choose the nontrivial value and a convenient local representative as
\begin{equation}
 S_{\mathrm{in}}=-\frac{1}{4\pi}\int_{N_4}F_Q\wedge F_F,
 \qquad \partial N_4=M_3.
 \label{eq:Sin}
\end{equation}
On closed manifolds $\exp(iS_{\mathrm{in}})=(-1)^{\int I_4}$.
The representatives $+\pi$ and $-\pi$ define the same closed-manifold phase but differ by an integer boundary counterterm when a boundary is present. The sign in Eq.~\eqref{eq:Sin} fixes the boundary-response convention used throughout this draft~\cite{Wang:2017txt}.
\subsection{Fractional mutual Chern-Simons level}
The boundary descendant of the inflow action can be represented locally by
the mutual Chern-Simons functional
\begin{equation}
 W_{\mathrm{loc}}=\frac12\C_{ab}\int_{M_3}A_a\wedge\dd A_b,
 \qquad {\rm where} \qquad
 \C_{ab}=\C_\star
 \begin{pmatrix}0&1\\1&0\end{pmatrix}_{ab},
 \qquad \C_\star=\frac{1}{4\pi}.
 \label{eq:localCS}
\end{equation}
On a closed patch-wise-trivial background this is $\C_\star\int A_Q\wedge\dd A_F$ up to a total derivative. In the normalization 
\beq
W=\frac{1}{4\pi} \int K_{ab}A_a\wedge\dd A_b ,
\eeq
the two off-diagonal entries are
\begin{equation}
 K_{QF}=K_{FQ}=\frac12 .
 \label{eq:KQF}
\end{equation}
The anomaly is given by a fractional mutual
Chern-Simons level $k_{QF}=\frac12 \,\,(\mathrm{mod}\ \mathbb Z)$. The fractional expression is a local representative of relative boundary response. By itself, it is not a globally defined intrinsic action for arbitrary compact bundles. An intrinsic mutual counterterm reads
\begin{equation}
 W_{\mathrm{ct}}=\frac{m}{2\pi}\int A_Q\wedge\dd A_F,
 \qquad m\in\mathbb Z.
 \label{eq:counterterm}
\end{equation}
Of course, changing its four-dimensional extension changes the action by $2\pi m\int I_4$. Thus $k=2\pi\C_\star$ is defined modulo integers, and the anomaly is represented by $k=1/2\pmod{\mathbb Z}$. Both $\mathsf C$ and $\mathsf P$ send the mutual response at level $k$ to that at level $-k$~\cite{Wang:2017txt}. Exact invariance of the local representative for arbitrary continuously variable sources would require $k=0$. Under $\CP$, however, the two signs cancel each other. This establishes the shared obstruction and its local $\CP$ invariance.

The compact charge lattice cannot be left out in this argument. Although real linear transformations of $(A_Q,A_F)$ are permissible for local algebra, a change of basis that preserves the specified unit-charge lattice lies in $GL(2,\mathbb Z)$. In particular, diagonalizing the response with $A_Q\pm A_F$ does not produce two independently normalized unit-charge connections without also changing the flux constraints.
\subsection{Global charge-conjugation backgrounds and a thermal test}
With a nontrivial $\mathsf C$ background, $A_Q$ is replaced by an $O(2)$ bundle $E_Q$. A global specification of the chosen bulk phase looks like
\begin{equation}
 Z_{\mathrm{in}}[X_4;E_Q,A_F] =(-1)^{\langle w_2(E_Q)\smile[c_1(F)]_2,[X_4]_2\rangle}.
 \label{eq:globalphase}
\end{equation}
The brackets denote mod-two reduction and evaluation on the mod-two fundamental class. For an oriented $E_Q$, $w_2(E_Q)=[c_1(Q)]_2$, reproducing Eq.~\eqref{eq:Sin} on closed oriented manifolds. Equation~\eqref{eq:globalphase} specifies this bulk sector and its mod-two form allows evaluation without a spacetime orientation. The local differential-form treatment below is restricted to oriented
patches with trivialized $\mathsf C$ background.

For a nontrivial evaluation, assume $X_4=T^2\times\Sigma_2$, one unit of $Q$ flux through $T^2$, and one unit of $F$ flux through $\Sigma_2$. Then $\int I_4=1$ and $Z_{\mathrm{in}}=-1$. One cycle of $T^2$ can be the thermal circle, with the other implementing a unit-winding large $Q$ gauge transformation. This probes the global inflow phase on a thermal mapping torus. It must not be interpreted as a sign change of a fully gauge-invariant partition function without providing its anomaly trivialization and regulator.
\section{Hydrostatic functional and transgression}
\label{sec:hydrostatic}
In equilibrium, there is a timelike thermal Killing vector field $\beta^\mu=u^\mu/T$, which is the ratio of the fluid velocity to the temperature. Gauge-covariant stationarity means $\mathcal L_\beta A_a+\dd\Lambda_{\beta a}=0$, with
\begin{equation}
 \frac{\mu_a}{T}=\beta^\mu A_{a\mu}+\Lambda_{\beta a}.
 \label{eq:thermalpotential}
\end{equation}
The relation $\mu_a=u^\mu A_{a\mu}$ is used only in a stationary gauge with $\Lambda_{\beta a}=0$. Writing $a_\mu=u^\nu\nabla_\nu u_\mu$ implies
\begin{equation}
 \nabla_\mu T=-Ta_\mu,
 \qquad \nabla_\mu\mu_a=E_{a\mu}-\mu_a a_\mu,
 \qquad \nabla_\mu u_\nu=-u_\mu a_\nu+\omega_{\mu\nu}.
 \label{eq:equilibrium}
\end{equation}
To first-order in derivatives, the gauge-invariant hydrostatic functional reads
\begin{equation}
 W_{\mathrm{hs}}=\int_{M_3}\vol_3
 \left[p+\alpha_QB_Q+\alpha_FB_F+\alpha_\Omega\Omega\right]
 +O(\partial^2).
 \label{eq:Whs}
\end{equation}
This denotes the Lorentzian continuation of the stationary functional, not an action for dissipative time evolution. We assume that the regular hydrostatic sector preserves parity and has no parity-odd order parameter. Since the thermodynamic scalars are parity even, this excludes independent regular terms linear in $B_a$ or $\Omega$. If that assumption is relaxed, the coefficients below specify the anomaly contribution only.

Extending the equilibrium into $N_4$, we define the transverse connections as
\begin{equation}
 A_a^\perp=A_a+\mu_a u,
 \qquad
 A_a(\tau)=A_a+(1-\tau)\mu_a u,
 \quad 0\leq\tau\leq1.
\end{equation}
Their stationary curvatures $F_a^\perp$ are horizontal with respect to the thermal direction. This leaves only three independent directions in $N_4$, $F_Q^\perp\wedge F_F^\perp=0$. Differentiating the product of the interpolating curvatures allows
\begin{align}
 F_Q\wedge F_F-F_Q^\perp\wedge F_F^\perp&=\dd V_{QF},\nonumber\\
 V_{QF}&=-\int_0^1\dd\tau\,
 \left[\mu_Q u\wedge F_F(\tau)+\mu_Fu\wedge F_Q(\tau)\right]\nonumber\\
 &=-\mu_Q u\wedge F_F-\mu_Fu\wedge F_Q
   -\mu_Q\mu_Fu\wedge\dd u\nonumber\\
 &=-\mu_Q u\wedge F_F-\mu_Fu\wedge F_Q
   -2\mu_Q\mu_Fu\wedge\omega.
 \label{eq:V}
\end{align}
Now the terms with $u\wedge\dd\mu_a\wedge u$ vanish. The last line uses $\dd u=-u\wedge a+2\omega$. Setting both connections and chemical potentials equal in Eq.~\eqref{eq:V} gives the full diagonal transgression as $-2\mu u\wedge F-2\mu^2u\wedge\omega$. The wedge-to-component identities following Eq.~\eqref{eq:conventions} give
\begin{equation}
 V_{QF}=(\mu_QB_F+\mu_FB_Q+\mu_Q\mu_F\Omega)\vol_3.
 \label{eq:Vcomponents}
\end{equation}
Thus Eq.~\eqref{eq:Sin} reduces to
$-\C_\star\int V_{QF}$. The cancellation of its local anomalous variation by the boundary functional permits only
\begin{equation}
 \alpha_Q=\C_\star\mu_F,\qquad
 \alpha_F=\C_\star\mu_Q,\qquad
 \alpha_\Omega=\C_\star\mu_Q\mu_F.
 \label{eq:alphas}
\end{equation}
Equivalently, one can write $\alpha_a=\C_{ab}\mu_b$ and $\alpha_\Omega=\C_{ab}\mu_a\mu_b/2$. Each scalar in Eq.~\eqref{eq:Vcomponents} is odd under $\mathsf C$ and $\mathsf P$, but even under their combined product. The vortical term is necessary even though the response is locally represented by a metric-independent Chern-Simons functional. In stationary coordinates, one can write
\begin{equation}
 ds^2=-e^{2\varphi}(dt+a_i\dd x^i)^2+h_{ij}\dd x^i\dd x^j,
 \quad A_a=A_{a0}(dt+a)+\widehat A_a,
 \quad \mu_a=e^{-\varphi}A_{a0}.
\end{equation}
Then, after discarding a spatial total derivative, we have
\begin{equation}
 \int V_{QF}=\int dt\wedge
 \left[A_{Q0}\dd\widehat A_F+A_{F0}\dd\widehat A_Q
+A_{Q0}A_{F0}\dd a\right]
 =\int A_Q\wedge\dd A_F.
 \label{eq:stationarycheck}
\end{equation}
This provides a direct stationary-coordinate check of both the coefficient and the sign of the mixed vortical term. One can use either this local Chern-Simons representation or Eq.~\eqref{eq:alphas}. However, adding them results in double counting of the anomaly response.
\section{Constitutive relations and entropy production}
\label{sec:constitutive}
\subsection{A thermodynamic-frame construction}
The local response in Eq.~\eqref{eq:localCS} provides a particularly simple off-equilibrium extension of the anomaly sector. Varying it gives
\begin{equation}
 j_{a,A}^\mu=\frac12\C_{ab}\epsilon^{\mu\nu\rho}F_{b\nu\rho}
 =\C_{ab}(B_bu^\mu+\widetilde E_b^\mu),
 \qquad T_A^{\mu\nu}=0.
 \label{eq:topcurrent}
\end{equation}
The stress variation is zero because the differential-form functional does not involve the metric. This statement concerns local bulk response. It does not specify possible physical edge stress tensors. The Bianchi identity and the symmetry of $\C$ imply
\begin{equation}
 \nabla_\mu j_{a,A}^\mu=0,
 \qquad \sum_aF_a^{\nu\lambda}j_{a,A\lambda}=0.
 \label{eq:topidentities}
\end{equation}
For the second identity, the contraction $F_a^{\nu\lambda}(\star F_b)_\lambda$ is antisymmetric in $a,b$, whereas $\C_{ab}$ is symmetric. This also tells why retaining both mutual currents is essential which means their Lorentz forces cancel as a pair.

Let a subscript $0$ denote the thermodynamic variables in this construction. We write
\begin{equation}
 V_a^\mu=E_a^\mu-T\Delta^{\mu\nu}\nabla_\nu(\mu_a/T),
 \quad \theta=\nabla_\mu u^\mu,
 \quad \sigma^{\mu\nu}=2\Delta^{\mu\alpha}\Delta^{\nu\beta}
 \nabla_{(\alpha}u_{\beta)}-\Delta^{\mu\nu}\theta.
\end{equation}
A minimal first-order constitutive relation for the stress-energy tensor and charge current can be written as
\begin{align}
 T^{\mu\nu}&=\epsilon_0u^\mu u^\nu+p_0\Delta^{\mu\nu}
 -\eta\sigma^{\mu\nu}-\zeta\theta\Delta^{\mu\nu}
 +O(\partial^2),\nonumber\\
 J_a^\mu&=n_{a0}u^\mu+\sigma_{ab}V_b^\mu+j_{a,A}^\mu
 +O(\partial^2).
 \label{eq:thermoframe}
\end{align}
For this minimal model we choose $\sigma_{ab}$ symmetric and positive semi-definite transport coefficients (shear and bulk viscosities), $\eta,\zeta\geq0$, and set optional non-hydrostatic Hall coefficients to zero. This close a consistent example, the inflow calculation alone does not determine every non-equilibrium coefficient. In particular, hydrostatic matching is not a derivation of a general real-time Hall conductivity. Equation~\eqref{eq:topidentities} reduce the interior Ward identities to those for the normal fluid. An entropy current is therefore written as
\begin{equation}
 S^\mu=s_0u^\mu-\frac{\mu_{a0}}{T_0}\sigma_{ab}V_b^\mu,
\end{equation}
with
\begin{equation}
 \nabla_\mu S^\mu=\frac{\eta}{2T}\sigma_{\mu\nu}\sigma^{\mu\nu}
 +\frac{\zeta}{T}\theta^2
 +\frac1T\sigma_{ab}V_a^\mu V_{b\mu}+O(\partial^3).
 \label{eq:entropyproduction}
\end{equation}
\subsection{Matching both charge densities in the Landau frame}
For the Landau variables, we impose the definitions
\begin{equation}
 u_\nu T^{\mu\nu}=-\epsilon u^\mu,
 \qquad -u_\mu J_a^\mu=n_a(T,\mu_Q,\mu_F).
 \label{eq:Landau}
\end{equation}
The fluid velocity already meets the energy condition in Eq.~\eqref{eq:thermoframe}. Only scalar matching is required for this anomaly sector. The transformation from the old thermodynamic variables to the new Landau variables is defined without a sign ambiguity using
\begin{equation}
 \epsilon(T,\mu_a)=\epsilon_0,
 \qquad n_a(T,\mu_b)=n_{a0}+\C_{ab}B_b.
 \label{eq:matching}
\end{equation}
In particular, if $x=(T,\mu_Q,\mu_F)$ and
$\delta x=x_{\mathrm L}-x_0$, the susceptibility matrix follows
\begin{equation}
 \frac{\partial(\epsilon,n_Q,n_F)}{\partial(T,\mu_Q,\mu_F)}\delta x
 =\begin{pmatrix}0\\\C_\star B_F\\\C_\star B_Q\end{pmatrix}.
 \label{eq:framechange}
\end{equation}
All quantities on the LHS can be computed at the zeroth order. We then write
\begin{equation}
 p_a=\left(\frac{\partial p}{\partial n_a}\right)_{
 \epsilon,n_{b\ne a}}.
\end{equation}
Since the physical isotropic stress is $p_0$, Eq.~\eqref{eq:matching} results in (to first-order)
\begin{equation}
 p_0=p(\epsilon,n_a)-p_a\C_{ab}B_b+O(\partial^2).
\end{equation}
Subsequently, the complete first-order Landau-frame relations for the energy-momentum tensor and the charge currents become:
\begin{align}
 T^{\mu\nu}&=\epsilon u^\mu u^\nu+
 \left[p-\zeta\theta-\C_\star(p_FB_Q+p_QB_F)\right]\Delta^{\mu\nu}
 -\eta\sigma^{\mu\nu},\nonumber\\
 J_Q^\mu&=n_Qu^\mu+\sigma_{Qb}V_b^\mu+\C_\star\widetilde E_F^\mu,\nonumber\\
 J_F^\mu&=n_Fu^\mu+\sigma_{Fb}V_b^\mu+\C_\star\widetilde E_Q^\mu.
 \label{eq:finalconstitutive}
\end{align}
Note that the pressure correction is a consequence of density matching only. Defining $\chi_{B_b}$ as the
\emph{additive} coefficient of $B_b\Delta^{\mu\nu}$, and
$\chi^a{}_{E_b}$ as that of $\widetilde E_b^\mu$, the anomaly contributions~\cite{Jensen:2011xb} are written as
\begin{equation}
 \chi_{B_Q}=-\C_\star p_F,\quad
 \chi_{B_F}=-\C_\star p_Q,\quad
 \chi^Q{}_{E_F}=\chi^F{}_{E_Q}=\C_\star,\quad
 \chi_\Omega=\chi_T^Q=\chi_T^F=0.
 \label{eq:coefficients}
\end{equation}
The diagonal electric coefficients vanish in this chosen mixed sector. If the stress is instead written with $-\widetilde\chi_B B\Delta^{\mu\nu}$, then $\widetilde\chi_B=-\chi_B$.

This derivation checks the hydrostatic result independently: the full local current and vanishing stress follow from the direct variation of the contact functional, after which Eq.~\eqref{eq:coefficients} follows only from the first law and frame matching. In the Landau frame, the same entropy current can be written as
\begin{equation}
 S^\mu=s u^\mu-\frac{\mu_a}{T}\nu_a^\mu
+\frac{\mu_a}{T}j_{a,A}^\mu+O(\partial^2),
 \qquad \nu_a^\mu=J_a^\mu-n_au^\mu.
 \label{eq:Landauentropy}
\end{equation}
Indeed, it can be seen that $s-s_0=-\mu_a\C_{ab}B_b/T$ at the fixed energy, so that the longitudinal terms in Eq.~\eqref{eq:Landauentropy} reproduce $s_0u^\mu$.

A general remark is as follows. The equations of state, viscosities, and the dissipative conductivity matrix remain independent constitutive information. A more general non-equilibrium model may add a Hall-viscosity term and transverse terms $\widetilde\sigma_{ab}\widetilde V_b^\mu$, all of which vanish in hydrostatic equilibrium. For multiple charges, however, the entropy test must retain the charge indices: $V_a\cdot\widetilde V_b=-V_b\cdot\widetilde V_a$. Thus, the charge-symmetric part of $\widetilde\sigma_{ab}$ is non-dissipative. Its charge-antisymmetric part does not automatically drop out of entropy production, and positivity must be imposed on the full conductivity operator in charge and spatial indices. The minimal choice in Eq.~\eqref{eq:thermoframe} avoids this complexity. First-order dissipative constitutive relations are used here only as a long-wavelength expansion.
\section{Static response and one analytic example}
\label{sec:response}
\subsection{Magnetization and contact response}
In a static flat background with $u^\mu=(1,0,0)$, the equilibrium magnetizations obtained from Eq.~\eqref{eq:Whs} are
\begin{equation}
 M_Q=\C_\star\mu_F,\qquad M_F=\C_\star\mu_Q.
\end{equation}
When the temperature is fixed and spatial magnetic fields are vanishing, we have
\begin{equation}
 J_{a,A}^i=\C_{ab}\epsilon^{ij}\partial_j\mu_b,
 \qquad E_{bi}=\partial_i\mu_b.
 \label{eq:equilibriumcurrent}
\end{equation}
These are divergence-free equilibrium magnetization currents. The reciprocal local magnetic-density response is written as
\begin{equation}
 \left.\frac{\partial(-u\cdot J_Q)}{\partial B_F}\right|_{T_0,\mu_{a0}}
 =\left.\frac{\partial(-u\cdot J_F)}{\partial B_Q}\right|_{T_0,\mu_{a0}}
 =\C_\star.
 \label{eq:Streda}
\end{equation}
Note that the derivatives hold the physical equilibrium sources fixed. Restoring probe charges gives the fractional cross coefficient $e_Qe_F/(2h)$, modulo $e_Qe_F/h$. To fix the signs, we define the static source-response kernel as
\begin{equation}
 \mathcal K_{ab}^{i0}(0,\mathbf k)
 =\left.\frac{\delta\langle J_a^i\rangle}
 {\delta A_{b0}}\right|_{\omega=0}.
\end{equation}
For the anomaly contribution, we have
\begin{equation}
 \mathcal K_{ab,A}^{i0}=i\C_{ab}\epsilon^{ij}k_j,
 \qquad {\rm where} \qquad
 \C_{ab}=\lim_{k_x\to0}\frac{i}{k_x}
\mathcal K_{ab,A}^{y0}(0,k_x).
 \label{eq:staticKubo}
\end{equation}
In particular, $\mathcal K^{y0}_{ab,A}=-i\C_{ab}k_x$. Identifying this kernel with a retarded correlator requires a specified source coupling and inclusion of its local contact terms~\cite{Closset:2012vp}. Equation~\eqref{eq:staticKubo} sets $\omega=0$ before $\mathbf k\to0$. For a homogeneous perturbation $A_{by}(t)$, instead, $E_{by}=i\omega A_{by}$ and the minimal local sector gives $\mathcal K^{xy}_{ab,A}(\omega,0)=i\omega\C_{ab}$. A homogeneous conductivity also samples optional non-hydrostatic terms, momentum dynamics, and the chosen contact and magnetization description. It is therefore the fractional local equilibrium coefficient, rather than an arbitrary full DC conductivity, that is characterized by the inflow.

If time reversal is imposed with density parities $\eta_Q=+1$ and $\eta_F=-1$, its action reverses the equilibrium parameter $\mu_F$. Microscopic reciprocity allows the Hall-sector relation to be written as
\begin{equation}
 \sigma^H_{ab}(T,\mu_Q,\mu_F)
 =-\eta_a\eta_b\sigma^H_{ba}(T,\mu_Q,-\mu_F)
 \label{eq:Onsager}
\end{equation}
at zero time-reversal odd sources. Equal-state reciprocity obeys at $\mu_F=0$, which means cross Hall entries are symmetric while diagonal entries vanish. When $\mu_F$ is non-vanishing, diagonal entries are only constrained to be odd between the two time-reversed states, and the constant off-diagonal matrix in Eq.~\eqref{eq:localCS} satisfies the relation \eqref{eq:Onsager}.
\subsection{An analytic example}
In this section, we study an example. For which, we assume
\begin{equation}
 p=aT^3+\frac{T}{2}(b_Q\mu_Q^2+b_F\mu_F^2),
 \qquad {\text{where}} \qquad a,b_Q,b_F>0,
 \label{eq:exampleEOS}
\end{equation}
and restrict to a neighborhood of $\mu_Q=\mu_F=0$.
The thermodynamic functions are therefore
\begin{equation}
 n_a=b_aT\mu_a,\qquad
 s=3aT^2+\frac12(b_Q\mu_Q^2+b_F\mu_F^2),\qquad
 \epsilon=2p.
\end{equation}
The Hessian of $p$ in $(T,\mu_Q,\mu_F)$ is positive definite when $b_Q\mu_Q^2+b_F\mu_F^2<6aT^2$, providing a sufficient and necessary local stability domain. Since $\epsilon=2p$, $p_Q=p_F=0$, meaning the magnetic pressure corrections vanish, although the cross-charge Hall response is non-vanishing.

Take constant transport coefficients evaluated at a neutral equilibrium, i.e., $T=T_0$ and $\mu_a=0$ with $\sigma_{ab}=\sigma_a\delta_{ab}$, $\sigma_a>0$, $\eta>0$, and $\zeta=0$. This allows the charge fluctuations to decouple from energy and momentum to linear order. With the vanishing external field, we receive
\begin{equation}
 \partial_t\delta n_a-D_a\nabla^2\delta n_a=0,
 \qquad \text{with} \qquad D_a=\frac{\sigma_a}{b_aT_0},
 \qquad \text{and} \qquad \omega_a=-iD_a k^2.
 \label{eq:diffusion}
\end{equation}
For a weak scalar source $\delta A_{a0}$ and vanishing magnetic field, the linearized current and density equation simplifies
\begin{align}
 \delta J_a^i&=\sigma_a\partial_i(\delta A_{a0}-\delta\mu_a)
+\C_{ab}\epsilon^{ij}\partial_j\delta A_{b0},\nonumber\\
 (-i\omega+D_a k^2)\delta n_a&=\sigma_a k^2\delta A_{a0}.
 \label{eq:drivenexample}
\end{align}
Here, the cross response is transverse in nature, and its divergence identically reduces to zero. It therefore changes the source response without adding a pole to the charge-density propagator. In particular, at $k_y=0$, we have $\delta J_Q^y=-i\C_\star k_x\delta A_{F0}$ even though the two diffusive denominators remain decoupled.

At the vanishing source, the shear and sound modes are
\begin{equation}
 \omega_{\mathrm{sh}}=-i\frac{\eta}{w_0}k^2,
 \qquad
 \omega_\pm=\pm\frac{k}{\sqrt2}
-i\frac{\eta}{2w_0}k^2+O(k^3),
 \qquad w_0 = 3 \, a\, T_0^3.
 \label{eq:sound}
\end{equation}
These expressions stem from the longitudinal determinant $$\omega^2-k^2/2+i(\eta/w_0)\omega k^2=0$$ and the transverse momentum equation. Together with Eq.~\eqref{eq:diffusion}, they provide the five hydrodynamic modes of this two-charge example. Their non-dependence on $\C_\star$ is a direct result of the source-local topological current.
\section{Physical edges and the limits of bulk matching}
\label{sec:edges}
A spatial edge of the fluid is separate from the auxiliary bulk boundary, $M_3=\partial N_4$. For a fluid region $\Sigma$, let's say $\Theta_\Sigma$ is its characteristic function. Even though the current in Eq.~\eqref{eq:topcurrent} is identically conserved in the interior
\begin{equation}
 \nabla_\mu(\Theta_\Sigma j_{a,A}^\mu) =(\nabla_\mu\Theta_\Sigma)j_{a,A}^\mu,
 \label{eq:edgeleakage}
\end{equation}
an edge completion or specified exchange with the exterior is therefore required. Bulk inflow does not imply arbitrary boundary conditions or all dynamical degrees of freedom at the edge.

For an impermeable wall, $u^\mu\nabla_\mu\Theta_\Sigma=0$, we define $\ell^\mu=\epsilon^{\mu\nu\rho}(\nabla_\nu\Theta_\Sigma)u_\rho$. Within the tangential edge-current ansatz, the known multi-charge result of Ref.~\cite{Furusawa:2021vlw}, Eq.~(37), is
\begin{align}
 J_{a,\mathrm e}^\mu&=\left[-\C_{ab}\mu_b+
\frac{n_a}{w}(H+cT^2)\right]\ell^\mu,
 & H&=\frac12\C_{ab}\mu_a\mu_b=\C_\star\mu_Q\mu_F,
 \nonumber\\
 S_{\mathrm e}^\mu&=-\frac{\mu_a}{T}J_{a,\mathrm e}^\mu
 -\left(\frac{H}{T}+cT\right)\ell^\mu.
 \label{eq:edgecurrents}
\end{align}
They assume a boundary sector described by this ansatz and leave the constant $c$ undetermined by the gauge anomaly.

For this specific symmetry, there is a useful extra information. The scalar $H$ is $\mathsf C$ odd, whereas $T^2$ is even. The vector $\ell^\mu$ is a parity pseudo-vector. If the edge geometry, boundary condition, and constitutive sector preserve $\CP$, the $H$ contribution has the required transformation, while an independent constant $cT^2\ell^\mu$ energy-flow contribution is $\CP$ odd. For that restricted edge sector one must have $c=0$. We remind the reader that this is a boundary-symmetry restriction and not a determination of $c$ from the mixed gauge coefficient. A boundary that breaks $\CP$ symmetry, or has further edge sectors, does not follow this restriction merely because the bulk has the given anomaly.

The same distinction applies to the use of the local contact functional on a region with a boundary. On a closed $M_3$, the symmetric and un-symmetrized mutual Chern-Simons expressions differ only by a total derivative. At a physical edge, that derivative changes the allocation of contact currents between the interior and the edge. The chosen current convention, edge action, and inflow must be treated together.
\section{Summary}
\label{sec:summary}
We have worked out a first-order (in gradients) hydrodynamic formulation of a mixed discrete anomaly in a relativistic fluid consisting of two compact $U(1)$ charges in $2+1$D. Charge conjugation reverses one charge and leaves the other charge invariant, while parity acts through spatial reflection. For this symmetry setup, the fractional mutual Chern-Simons response represents a shared obstruction to preserving charge conjugation and parity separately, with their product preserved in the chosen anomaly sector. The compact charge lattice is essential to this statement, which means that admissible intrinsic counterterms shift the mutual level by integers and cannot remove its half-integer part. 

Making use of the previously constructed multi-charge anomaly sector, we followed its consequences through hydrostatic matching, constitutive relations, and static response. Transgression determines the coupled magnetic and vortical terms in the equilibrium functional. Moreover, the direct variation of the local contact functional provides an independent derivation of the anomalous currents and shows that their local bulk stress contribution is vanishing. The currents are identically conserved in the interior, and their combined Lorentz force is zero. Consequently, the anomaly can be incorporated without modifying the local charge Ward identities or introducing additional bulk entropy production. Matching both charge densities at fixed energy produces magnetic pressure corrections in the Landau frame. These corrections arise from the thermodynamic redefinition rather than from independent topological stress.

The response analysis identifies where this anomaly is present and what its coefficient determines. It gives reciprocal magnetic density responses and a transverse equilibrium contact response whose fractional part is fixed by the anomaly class. Their interpretation requires a specified current convention and order of limits. In particular, equilibrium magnetization currents do not by themselves determine transport between reservoirs, and hydrostatic matching does not specify a general homogeneous DC conductivity. The analytic neutral-fluid example, we presented, makes a related distinction manifest. The mixed transverse source response is non-vanishing, although the source free sound, shear, and charge diffusion poles remain independent of the anomaly coefficient.

Our model distinguishes the fractional anomaly response from the thermodynamic and dissipative information that remains unconstrained. 
\begin{acknowledgments}
RS acknowledges the support of the Humboldt research fellowship for postdocs from the Alexander von Humboldt Foundation and the kind hospitality of the Department of Physics, IIT Guwahati. SP acknowledges financial support from the IIT Guwahati Postdoctoral Fellowship under Project No. IITG/AR/IPDF/2025-27/007. We acknowledge the use of AI tools (Claude \& OpenAI) for cross-checking the calculations and for proof-reading the manuscript. The authors take full responsibility for the content of the manuscript.
\end{acknowledgments}
\bibliographystyle{utphys}
\bibliography{pv_ref}
\end{document}